\documentstyle[12pt,aas2pp4,epsfig]{article}

\def\be{\begin{equation}}
\def\ee{\end{equation}}
\def\ea{{\it et al.}\,}
\def\np{\newpage}
\def\eg{{\it e.g.},\,}

\def\rel{relativistic \,}

\begin{document}

\hfill\today
\title{RXTE Observations of A2256}

\author{Yoel Rephaeli\altaffilmark{1,2}, and Duane 
Gruber\altaffilmark{3}}

\affil{$^1$Center for Astrophysics and Space Sciences, 
University  of California, San Diego,  La Jolla, CA\,92093-0424}

\affil{$^2$School of Physics and Astronomy, 
Tel Aviv University, Tel Aviv, 69978, Israel}

\affil{$^3$4789 Panorama Drive, San Diego CA 92116}

\begin{abstract}
The cluster of galaxies A2256 was observed by the PCA and HEXTE 
experiments aboard the RXTE satellite during the period July 2001 - 
January 2002, for a total of $\sim$343\,ks and $\sim$88\,ks, 
respectively. Most of the emission is thermal, but the data analysis 
yields evidence for two components in the spectrum. Based on 
statistical likelihood alone, the secondary component can be either 
thermal or power-law. Inclusion in the analysis of data from ASCA 
measurements leads to a more definite need for a second component. 
Joint analysis of the combined RXTE-ASCA data sets yields $kT_1 = 
7.9^{+0.5}_{-0.2}$ and $kT_2 = 1.5^{+1.0}_{-0.4}$, when the second 
component is also thermal, and $kT = 7.7^{+0.3}_{-0.4}$ and $\alpha = 
2.2^{+0.9}_{-0.3}$, if the second component is fit by a power-law with 
(photon) index $\alpha$; all errors are at 90\% confidence. Given the 
observed extended regions of radio emission in A2256, it is reasonable 
to interpret the deduced power-law secondary emission as due to Compton 
scattering of the radio producing relativistic electrons by the cosmic 
microwave background radiation. If so, then the {\it effective, mean 
volume-averaged} value of the magnetic field in the central 1$^{o}$ 
region of the cluster -- which contains both the `halo' and `relic' 
radio sources -- is $B \sim 0.2^{+1.0}_{-0.1}$ $\mu G$.
\end{abstract}


\section{Introduction} 

The improved spatial resolution and wider spectral coverage 
of current X-ray satellites provide further motivation for 
a less simplified description of the properties of intracluster 
(IC) gas, and the consideration of additional cluster phenomena,  
such as non-thermal (NT) processes. An obvious generalization of 
the simple isothermal model for the gas is made by allowing for a 
more realistic temperature structure. This has, in fact, been 
considered in quite a few analyses of cluster X-ray data, leading 
to clear evidence for radial variation of the gas temperature in 
some clusters (\eg, Honda \ea 1996, Watanabe \ea 1999, Markevitch 
\ea 1998). Somewhat less obvious is the need to include a NT 
component in the X-ray spectra of (at least some) clusters. Such 
emission has long been predicted (\eg Rephaeli 1977; for a review, 
see Rephaeli 2001), and recent observations give appreciable evidence 
for its likely detection in a few clusters.

It is of considerable interest to know if IC gas within the 
central ($\sim 1$ Mpc) cluster region is non-isothermal. In addition 
to insight gained from the form of the temperature profile on 
physical processes in the gas, and its cosmological evolution, 
knowledge of the density and temperature distributions is clearly 
very important not only for the determination of basic cluster 
properties, such as the gas and dark matter masses, but also for 
the use of phenomena in clusters to determine cosmological parameters 
(\eg, from measurements of the Sunyaev-Zeldovich effect and 
gravitational lensing). On the other hand, measurement of a NT 
spectral component -- especially in clusters in which extended 
regions of radio emission have been measured --  is essential for 
the characterization of NT quantities and phenomena in clusters, 
such as the strength and morphology of magnetic fields, densities 
and energy content of \rel electrons and protons, and the interaction 
of these particles with the gas.

Measurements with RXTE and BeppoSAX satellites are particularly 
useful in the search for NT spectral components in cluster spectra.
First attempts to detect NT emission from a few clusters 
with the HEAO-1, CGRO, and ASCA satellites were unsuccessful (Rephaeli, 
Gruber \& Rothschild 1987, Rephaeli \& Gruber 1988, Rephaeli, Ulmer 
\& Gruber 1994, Henriksen 1998). The improved sensitivity and wide 
spectral band of the RXTE and BeppoSAX seem to have resulted in the 
detections of NT emission in Coma (Rephaeli, Gruber \& Blanco 1999, 
Fusco-Femiano \ea 1999, Rephaeli \& Gruber 2002), and A2319 
(Gruber \& Rephaeli 2002). A NT component was also deduced in the 
BeppoSAX spectra of A2256, A119, and A754 (Fusco-Femiano \ea 2002). 
Here we report the results from a long RXTE observation of A2256, 
possibly a merging cluster with complex X-ray (Sun \ea 2002) and 
radio (Giovannini, G., \ea 1999) morphologies.

\section{Observations and Data Reduction}

A2256 is a rich radio and X-ray bright cluster at $z=0.0581$, with 
complex morphology. Extended radio emission from A2256 was measured 
by Bridle \& Fomalont (1976), Bridle \ea (1979), Giovannini \ea
(1999), and most recently by Clarke \& Ensslin (2001). In addition to 
emission from several strong radio sources in the central region of 
the cluster, there is a centrally located extended emission region, 
as well as regions of extended emission (located in the northern side 
of the cluster) that are thought to be radio relics. The centrally 
located emission is characterized by a spectral (energy) index 
$\alpha_C \sim 2.0$ in the $\sim 1.4 - 5$ GHz band, while the main 
relic has a flatter spectrum with index $\alpha_R \sim 1.0$ (Clarke 
\& Ensslin (2001).   

The cluster was observed by most previous X-ray satellites, most 
recently by XMM and {\it Chandra}. Several emission regions were 
resolved by {\it Chandra}, yielding further evidence for the view 
that the cluster is undergoing a merger (Sun \ea 2002). These 
measurements also indicate considerable variation of the temperature 
in the central region, with a mean value of $\simeq 6.7$ keV, but 
with hotter ($\sim 10$ keV) and colder ($\sim 5$ keV) regions. Of 
particular interest to us here are BeppoSAX observations with the MECS 
and PDS detectors: From an analysis of these measurements 
Fusco-Femiano \ea (2000) deduced the presence of a NT spectral 
component at a $\sim 4.6\sigma$ confidence, with a (photon) index 
roughly in the range $0.3 - 1.7$, in addition to the main thermal 
component with a temperature of $7.4 \pm 0.2$ keV. A short, $\sim 30$
ks, observation of A2256 by RXTE yielded only an upper limit on a NT 
flux, $\sim 2.3\times 10^{-7}$ cm$^{-2}\,s^{-1}\,keV^{-1}$ at 30 keV, 
and a lower limit of $\sim 0.36\, \mu$G on the mean magnetic field 
(Henriksen 1999).

A2256 was observed with RXTE for a nominal total observation
time of 400 ks between July 2001 and January 2002. After the
application of data selection criteria recommended by the RXTE
project, 343 ks of screened data were collected with the PCA in
111 one-orbit observations, spaced irregularly over the seven-month
campaign. For the HEXTE, which beam-switches observations with 32-second 
dwells between source and background fields, and has in addition about
50\% detector dead time, the net observation time was 88 ks with
each of the two clusters. On time scales of two weeks or longer,
the limit to variability observed with PCA was less than 1\%.
Because of the much lower signal to background, corresponding
HEXTE limits to variability are larger, about 20\%.

Following project practice at the time, PCA data were collected in two
of the 5 detectors.  One of these, PCU 0, had lost its propane guard
layer by the time of the A2256 campaign, but the net flux and spectral
shape differed negligibly from those obtained with PCU 2, which indicates
that the project has very successfully produced modifications for the
response matrix and background estimation tool for PCU 0.

ASCA observations of A2256 have been archived for several
observations, the longest of which was 36 ks on 22 July 1993, and
the next-longest, 26 ks, was carried out on 8 April 1993. Standard
archival GIS and SIS spectra and matrices were provided. Preliminary
spectral study of the two observations gave very similar best fits
of an isothermal spectral model. However, there was large 
non-statistical scatter of about 10\% in the SIS data of the longer 
July observation. For this reason the July data were considered less 
reliable, and the April data alone were employed in the joint analysis 
with the RXTE data.

\section{Spectral Analysis}

The combined ASCA and RXTE data provide spectral information on the 
rather broad energy range of 0.6 to 100 keV. The four ASCA detectors, 
two PCA detectors and two HEXTE clusters, with thousands of energy 
channels combined, rather heavily oversample this range. After pilot 
spectral studies revealed no evidence for sharp features in the raw 
data, we proceeded to reduce the spectral oversampling to a reasonable 
level by combining counts from similar detectors and by summing counts 
in adjacent spectral channels into groups whose energy width was set 
at about one half of the FWHM detector energy resolution at the given 
energy.  Appropriate response matrices were also generated, and 
standard ftools were employed. The final spectral set contained 44 
energy bands for the ASCA data and 44 for the RXTE data.

An additional systematic error of 0.5\% per energy channel was added 
in quadrature to the statistical error of the PCA data (e.g Wilms 
et al. 1999). No systematic error was used with HEXTE data, and 2\% 
systematic error was used for both the ASCA SIS and GIS data. 
Spectral analysis was performed separately on the RXTE data, jointly 
on the RXTE and ASCA data sets, and in a restricted analysis, also 
on just the HEXTE data above 15 keV. 

Spectral models were limited to three cases: an isothermal thermal 
spectrum (based on a Raymond-Smith emission code), two-temperature 
thermal, and a thermal plus a power-law. The RXTE data by themselves 
provide only weak evidence of the need for an extra component beyond 
isothermal. The $\chi^2$ of 46.8 (40 degrees of freedom [dof]) for an 
isothermal fit is acceptable; however inclusion of a second component 
reduces $\chi^2$ modestly to 40.2 (38 dof) with an extra 0.9 keV thermal 
component, or to 40.4 (38 dof) with an extra power-law, whose best-fit 
photon index is a rather steep 4.0. For four ``interesting'' parameters, 
the change in $\chi^2$ (Lampton \ea 1976) gives 90\% error limits for 
the 4-20 keV power-law flux of $(0.5-4.8)\times 10^{-11}$ 
erg-cm$^{-2}$ s$^{-1}$. The temperature for the main spectral component 
is in the rnage $\sim 7.7-7.9$ keV for all three cases, with formal 
($1\sigma$) errors of $\sim 0.1-0.3$ keV.

By fitting jointly with the ASCA data one obtains much more decisive
results.  An isothermal fit is ruled out both by a high $\chi^2$
of 155.4 for 82 dof, and by much improved fits with a second component,
$\chi^2$ = 96.9 with a second thermal at 1.4 keV, and $\chi^2$ = 104.5
with a power-law component with best-fit photon index 2.2.  For 
both of these cases the $\chi^2$ is somewhat high for 80 dof, but
this may reflect slight under-correction for systematic errors of
background subtraction and the response matrices.  With four 
interesting parameters (kT, abundance, power-law flux and index) 90\% 
error bounds for the power-law flux, now given for the interval 
0.8--40 keV, are $(2.5-19.1)\times 10^{-11}$ erg-cm$^{-2}$ s$^{-1}$.  
Best fit parameters and 90\% confidence errors for the joint fits to 
the three spectral models are listed in Table 1. (For each combination 
of detectors the energy range for the power-law flux has been chosen 
to provide parameter and error estimates which are nearly independent 
of the other parameters. This is approximately equal to the energy 
span of the joint data set.) In Figure 1, we show the spectrum of the 
best-fit isothermal plus power-law model (data and model components 
are displayed) in the upper panel, and residuals to the fit in the 
lower panel.

Most of the statistical weight in parameter estimation comes from
data at the lowest energies.  Of special interest for the thermal
plus power-law case is whether the HEXTE data favor the presence
of a power-law component. We tested the HEXTE data against a model
in which the thermal parameters are set by the joint fit. With
no second component the HEXTE data give a marginally acceptable
($P<0.06$) $\chi^2$ of 27.1 for 19 dof. When the power-law flux
is allowed to float to a best fit value the $\chi^2$ is dropped
by 9.3 to 17.8.  Allowing also the index to vary gives a best-fit
value of 1.8, which -- within errors -- is consistent with the value 
of 2.2 obtained in the joint fit. For one interesting parameter, 
the error bounds on the 15--40 keV flux are $(1.2-4.3)\times 10^{-12}$ 
erg-cm$^{-2}$ s$^{-1}$.

The NT 20-80 keV flux (of interest for a direct comparison with 
the BeppoSAX rasults) computed from the best-fit parameters resulting 
from the full (ASCA, PCA, and HEXTE) dataset is 
$(0.7-8.6)\times 10^{-12}$ erg-cm$^{-2}$ s$^{-1}$, formally 
significant at the 2.9 $\sigma$ level. The signficance of this 
flux is lower if HEXTE data are not included in the analysis. 
Note also that similar but less significant results are obtained 
when we first find the (poorly determined) best-fit isothermal to 
the PCA and ASCA data, and then determine a net high energy flux 
from the HEXTE data (a procedure adopted in the corresponding 
BeppoSAX MECS/PDS analysis). Doing so results in a 20-80 keV flux 
error bounds of $(0.2-8.0)\times 10^{-12}$ erg-cm$^{-2}$ s$^{-1}$, 
formally significant at 2.2 $\sigma$.

\begin{table*}        
\caption{Results of the spectral analysis}

\begin{tabular}{|l|ccc|} 
\hline 
Parameter & Single Thermal & Double Thermal & Thermal + Power-law \\                   
\hline   
$kT_1$ (\rm{keV}) &$7.66\pm 0.12$ & $7.91^{+0.48}_{-0.20}$ & $7.67^{+0.28}_{-0.39}$ \\      
                  &               &                        &                  \\
$kT_2$ (\rm{keV}) &               & $1.45^{+0.98}_{-0.35}$ &                  \\
                  &               &                        &                  \\            
$\alpha$          &               &                        &  $2.16^{+0.86}_{-.30}$ \\      
Secondary flux fraction &         &                        &                      \\       
\,\,\, 0.5-2 keV   &              & $0.084^{+0.069}_{-0.035}$  &                   \\
\,\,\, 2-10 keV    &              & $0.015^{+0.130}_{-0.060}$  &                   \\
\,\,\, 0.8-40 keV &             &                       & $0.101^{+0.083}_{-0.077}$ \\   
                  &               &                        &                  \\       
Abundance (solar)  & 0.194$\pm$0.018 & 0.208$\pm$0.028 & 0.218$\pm$0.030   \\                
\hline
\end{tabular} 
         
\tablenotetext{}{All quoted errors are at the 90\% confidence level. } 
\end{table*}               

Fusco-Femiano et al. (2000), reporting results of A2256 measurements 
with the BeppoSAX satellite, have claimed detection of a 20-80 keV 
NT flux of $1.2\times 10^{-11}$ ergs cm$^{-2}$ s$^{-1}$ at 4.6$\sigma$ 
significance. Assuming normal statistics we convert this 
to 90\% error bounds of [0.74, 1.66] in the same units.  Using the  
joint RXTE/ASCA dataset we obtain a comparable best-fit value for 
20-80 keV flux of 0.26 and 90\% confidence error bounds of 
[0.01, 0.79] in these units. Thus, while our best fit value is a 
factor of 4.6 smaller than that obtained by Fusco-Femiano et al. 
(2000), it is not in strong conflict. Fusco-Femiano have discussed 
the possibility that the radio and X-ray NT components are complex, 
with more than one index. In this case, our joint RXTE/ASCA NT flux 
may be sensitive largely to the steeper index visible to ASCA. Thus 
a better comparison may be with our HEXTE-only analysis. Indeed, 
this analysis corresponds rather closely to the approach adopted in 
the analysis of the BeppoSAX data. The HEXTE result gives a best-fit 
20-80 keV flux of $4.3 \cdot 10^{-12}$ ergs cm$^{-2}$ s$^{-1}$ and a 
90\% confidence error interval of $(0.3-10.0)\times 10^{-12}$ 
ergs cm$^{-2}$ s$^{-1}$. While smaller than the reported BeppoSAX value 
by almost a factor of three, this result is not in a great conflict. 

\section{Discussion}

A2256 is the third cluster with extended regions of radio emission 
that has been observed by the RXTE for more than 100 ks. The results 
of the analysis reported here are qualitatively similar to those we 
have previously reported on Coma (Rephaeli, Gruber \& Blanco 1999, 
Rephaeli \& Gruber 2002) and A2319 (Gruber \& Rephaeli 2002). In all 
three clusters the RXTE measurements yield evidence that the spectra 
in the combined PCA and HEXTE bands contain a secondary component which 
is either thermal or power-law. In the case of A2256 this evidence is 
much stronger when the ASCA data are included in the analysis. However, 
the spectral analysis alone does not yield sufficient statistical 
preference for the nature of the second component. We invoke other 
considerations in an attempt to determine the nature of this 
component.                     

Consider first thermal emission from IC gas at a lower temperature than 
that of the main emission component, as listed in Table 1. That deep 
obervations over a wide spectral range require a more realistic emission 
model than a single temperature gas is, of course, not unexpected. 
Indeed, recent mapping of the temperature in the central $\sim 1/2$ 
Mpc ($H_0$ = 70 km s$^-1$ Mpc$^-1$) radial region of A2256 by {\it 
Chandra} shows an emitting region (in the NE side of the cluster) 
with a fractional projected area of roughly $\sim 1/10$ where the gas 
temperature is $kT \sim 4$ keV. This value is still significantly higher 
than the value at the upper end of the 90\% significance interval, 
$kT_2 \simeq 2.4$ keV. Emission at this temperature with a fractional 
0.5-2 keV flux contribution of more than $\sim 5\%$ would have been 
measured, especially by ROSAT whose PSPC detector energy band matched 
this spectral range. Even so, we cannot formally rule out (also because 
of differences in collecting areas of the various detectors) the 
possibility that the second component is thermal at the above 
(relatively) low temperature. This is particularly so given the more 
realistic expectation that a two-temperature gas model is just a 
simplified representation of a more realistic continuous temperature 
distribution, as we have previously argued in the interpretation of 
RXTE measurements of Coma and A2319 (Rephaeli, Gruber \& Blanco 1999, 
Gruber \& Rephaeli 2002, Rephaeli \& Gruber 2002).
                                      
The main motivation for selecting A2256 for the long RXTE observation is, 
of course, the presence of extended regions of radio emission in the 
cluster. We first note that since significant emission from an AGN in the 
FOV is unlikely -- emission from QSO 4C +79, near the edge of the RXTE
FOV, was estimated to be negligible (Fusco-Femiano 2000) -- and given
no evidence for flux variability, we naturally associate the emission
with the cluster. As is well known (\eg, Rephaeli 1977), Compton
scattering of the radio producing relativistic electrons off the
cosmic microwave background boosts photon energies to the X-ray 
region. From the measured radio and X-ray fluxes the magnetic field can 
then be inferred. However, in the case of A2256 this is not 
straightforward because of the complex structure of the radio emission 
which is dominated by a few extended sources with spectral (energy) 
indices in the wide range, $\sim 0.3 - 1.1$, with substantial errors. 
Given no clear expectation on the predicted spectral index of the NT 
X-ray emission -- and the complex spatial morphology of the radio 
emission -- we use the measured total flux at 1.4 GHz, 397 mJy 
(Giovannini \ea 1999), and the deduced range of the X-ray power-law index 
to compute an {\it effective} value of the magnetic field across the 
large field of view of the RXTE, which includes all the dominant radio 
sources. Doing so, we determine the relatively wide 90\% confidence range 
for the {\it mean, volume-averaged} field, $B_{rx} \simeq 
0.2^{+1.0}_{-0.1} \,\mu$G. (The very high value at the 
upper end of this interval results from our conservatively estimated 
lower limit on the flux.) We emphasize that this value has only limited 
meaning: Since there is no spatial information on the power-law X-ray 
emission, the implicit assumption made here -- and in all similar 
analyses of cluster magnetic fields from radio and X-ray measurements -- 
is that these emissions occur over the same volume. If so, we can also 
{\it estimate} the mean \rel energy density within the emitting region, 
radius R, by integrating the electron energy distribution over 
energies in the observed radio and X-ray bands. Doing so, we obtain 
$\rho_{e} \simeq 5^{+1.0}_{-4}\times 10^{-14} (R/1 Mpc)^{-3}$ 
erg\,cm$^{-3}$.

Although the estimated mean field has limited meaning, it is comparable 
to the values we deduced for the mean field in Coma (Rephaeli \& Gruber 
2002) and A2319 (Gruber \& Rephaeli 2002). Values of the field deduced 
from radio and X-ray measurements, $B_{rx}$, are generally much lower 
than those obtained from Faraday rotation measurements (\eg, Clarke, 
Kronberg, and B\"ohringer 2001, and the review by Carilli \& Taylor 2002) 
of background radio sources seen through clusters, $B_{fr}$. The mean 
strength of IC fields has direct implications on the range of electron 
energies that are deduced from radio measurements, and therefore on the 
electron (Compton-synchrotron) loss time. The higher the electron energy, 
the shorter is the energy loss time; a short loss time would have 
immediate consequences on \rel electron models (\eg, Rephaeli 1979, 
Sarazin 1999, Ensslin \ea 1999, Brunetti \ea 2001, Petrosian 2001). 
Reliable estimates of the field are therefore quite essential.

Differences between $B_{rx}$ and $B_{fr}$ could, however, be due to the 
fact that the former is a volume-weighted measure of the field, whereas 
the latter is an average along the line of sight, weighted by the 
electron density. In addition, the field and \rel electron density would 
generally have different spatial profiles that could lead to very 
different spatial averages (Goldshmidt \& Rephaeli 1993). Various 
statistical and physical uncertainties in the Faraday rotation 
measurements, and their impact on deduced values of IC fields, were 
investigated recently by Newman, Newman \& Rephaeli (2002); their work 
strengthens the conclusion that a simple comparison of values of 
$B_{rx}$ and $B_{fr}$ is meaningless. More importantly, Rudnick \& 
Blundell (2003) have recently shown very clearly that the estimation 
of cluster fields from Faraday rotation measurements is very uncertain 
due to the inclusion in the sample of {\it cluster} radio sources whose 
large contributions to the rotation measures originate from their 
{\it intrinsic} fields, not the cluster-wide fields that they were 
presumed to sample.

RXTE and BeppoSAX measurements yielded evidence for NT X-ray 
emission in 5 clusters. It is important to continue the search 
for NT emission in other clusters with extended regions of radio 
emission. In particular, it is essential to obtain spatial information on 
this emission. This will likely be done for the first time by the IBIS 
instrument on the INTEGRAL satellite during a planned 500 ks observation 
of the Coma cluster. With the moderate $\sim 12'$ spatial resolution of 
IBIS, it should be feasible to determine the location of the region where 
the secondary emission is produced in this cluster.

\acknowledgments
We thank the referee, Dr. Roberto Fusco-Femiano, for his suggestions. 
This project has been supported by a NASA grant at UCSD.

\parskip=0.02in
\def\ref{\par\noindent\hangindent 20pt}

\np
\begin{figure}  
\centerline{\psfig{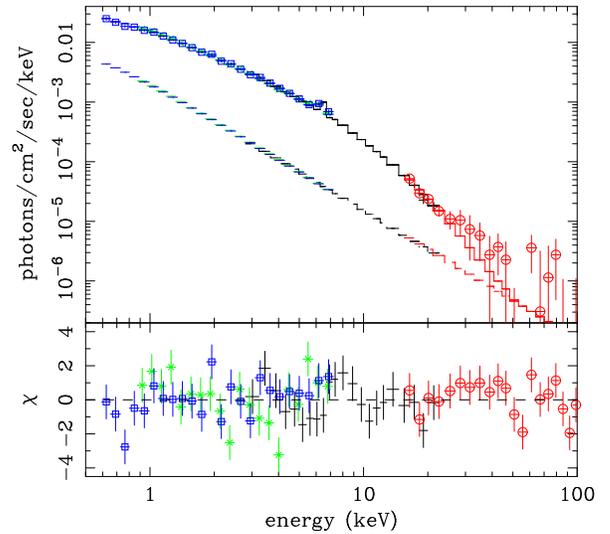}}
\figcaption{Joint RXTE-ASCA (photon) spectrum of the A2256 with folded 
Raymond-Smith ($kT \simeq 7.7$) and power-law (index $=2.2$) models. 
ASCA data are shown in green and blue circles; crosses are PCA data, and 
HEXTE data points are marked with red circles (with 68\% error bars). 
The total fitted spectrum is shown with a histogram, while the lower 
histogram shows the power-law portion of the best fit. The quality of 
the fit is demonstrated in the lower panel, which displays the observed 
difference normalized to the standard error of the data point.} 
\end{figure} 

\end{document}